\documentclass[pdflatex,sn-basic]{sn-jnl}
\usepackage{graphicx}
\usepackage{svg}
\usepackage{natbib}
\usepackage{csquotes}
\usepackage{booktabs}
\usepackage{rotating}
\usepackage{makecell}
 \usepackage{multirow}
\usepackage{diagbox}
\usepackage{array}
\newcolumntype{C}[1]{>{\centering\arraybackslash}p{#1}}
\usepackage{acro}
\DeclareAcronym{IS}{
short = IS,
long = information system
}

\DeclareAcronym{DSR}{
short = DSR,
long = design science research
}

\DeclareAcronym{BPM}{
short = BPM,
long = business process management
}

\DeclareAcronym{BAM}{
short = BAM,
long = business activity monitoring
}

\DeclareAcronym{BPMN}{
short = BPMN,
long = business process model and notation
}

\DeclareAcronym{RPM}{
short = RPM,
long = robotic process mining
}

\DeclareAcronym{EPC}{
short = EPC,
foreign-plural = {},
long = event-driven process chain
}

\DeclareAcronym{BN}{
short = BN,
long = Bayesian network
}

\DeclareAcronym{HMM}{
short = HMM,
long = hidden Markov model
}

\DeclareAcronym{DBN}{
short = DBN,
long = dynamic Bayesian network
}

\DeclareAcronym{DAG}{
short = DAG,
long = directed acyclic graph
}

\DeclareAcronym{2-TBN}{
short = 2-TBN,
long = 2-time-slice Bayesian network
}

\DeclareAcronym{EM}{
short = EM,
long = expectation maximization
}

\DeclareAcronym{PFA}{
short = PFA,
long = probabilistic finite automaton
}

\DeclareAcronym{BK}{
short = B\&K,
long = Boyen and Koller
}

\DeclareAcronym{NL}{
short = NL,
long = normalized likelihood
}

\DeclareAcronym{CECA-DBN}{
short = CECA-DBN,
long = cause-effect context-aware dynamic Bayesian network
}

\DeclareAcronym{PPM}{
short = PPM,
long = predictive process monitoring
}

\DeclareAcronym{PBPM}{
short = PBPM,
long = predictive business process monitoring
}

\DeclareAcronym{ML}{
short = ML,
long = machine learning
}

\DeclareAcronym{AI}{
short = AI,
long = artificial intelligence
}

\DeclareAcronym{CPD}{
short = CPD,
long = conditional probability distribution
}

\DeclareAcronym{ESA}{
short = ESA,
long = evidence sensitivity analysis
}

\DeclareAcronym{BPI}{
short = BPI,
long = business processing intelligence
}

\DeclareAcronym{BPIC}{
short = BPIC,
long = business processing intelligence challenge
}

\DeclareAcronym{RQ1}{
short = RQ1,
long = research question 1
}

\DeclareAcronym{RQ2}{
short = RQ1,
long = research question 2
}

\DeclareAcronym{XES}{
short = XES,
long = extensible event stream
}

\DeclareAcronym{MISQ}{
short = MISQ,
long = management information systems quarterly
}

\DeclareAcronym{ProM}{ 
short = ProM,
long = process mining workbench
}

\DeclareAcronym{HPC}{ 
short = HPC,
long = high performance computing
}

\DeclareAcronym{ADR}{ 
short = ADR,
long = action design research
}

\DeclareAcronym{SMR}{ 
short = SMRA,
long = super maximal repeat
}

\DeclareAcronym{SMRA}{ 
short = SMRA,
long = super maximal repeat alphabet
} 

\DeclareAcronym{CVI}{ 
short = CVI,
long = cluster validity indices
} 

\DeclareAcronym{ATR}{ 
short = ATR,
long = applied technology research
} 

\DeclareAcronym{CRISP-DM}{ 
short = CRISP-DM,
long = cross-industry standard process for data mining
}

\DeclareAcronym{BPMP-PM}{ 
short = BPMP-PM,
long = business process monitoring and prediction procedure model
}

\DeclareAcronym{IoT}{ 
short = IoT,
long = Internet of things
}

\DeclareAcronym{DNN}{ 
short = DNN,
long = deep neural network
}

\DeclareAcronym{RNN}{ 
short = RNN,
long = recurrent neural network
}

\DeclareAcronym{LSTM}{ 
short = LSTM,
long = long short-term memory neural network
}

\DeclareAcronym{CS-LSTM}{ 
short = CS-LSTM,
long = context sensitive LSTM
}

\DeclareAcronym{DL}{ 
short = DL,
long = deep learning
}

\DeclareAcronym{TP}{ 
short = TP,
long = true positive
}

\DeclareAcronym{FP}{ 
short = FP,
long = false positive
}

\DeclareAcronym{TN}{ 
short = TN,
long = true negative
}

\DeclareAcronym{FN}{ 
short = FN,
long = false negative
}

\DeclareAcronym{XAI}{ 
short = XAI,
long = explainable artificial intelligence
}

\DeclareAcronym{GNN}{ 
short = GNN,
long = graph neural network
}

\DeclareAcronym{GGNN}{ 
short = GGNN,
long = gated graph neural network
}

\DeclareAcronym{PDP}{ 
short = PDP,
long = partial dependence plot
}

\DeclareAcronym{LRP}{ 
short = LRP,
long = layer-wise relevance propagation
}

\DeclareAcronym{LIME}{
    short = LIME,
    long = local interpretable model-agnostic explanation
}

\DeclareAcronym{SHAP}{
    short = SHAP,
    long = Shapley additive explanation
}

\DeclareAcronym{DSS}{
    short = DSS,
    long = decision support system
}

\DeclareAcronym{RF}{
    short = RF,
    long = random forest
}
\DeclareAcronym{XPBPM}{
short = XPBPM,
long = explainable predictive business process monitoring
}

\DeclareAcronym{BPR}{
short = BPR,
long = business process redesign
}

\DeclareAcronym{RCA}{
short = RCA,
long = root cause analysis
}
\DeclareAcronym{PW}{
short = PW,
long = process warehousing
}

\usepackage{xcolor}%
\usepackage{textcomp}%
\usepackage{manyfoot}%
\usepackage{listings}%

\usepackage{amsthm,amsmath}%
\theoremstyle{thmstyleone}
\newtheorem{rqt}{RQ-T}
\newtheorem{rqd}{RQ-D}
\newtheorem{rqpk}{RQ-PK}
\newtheorem{rqp}{RQ-P}
\newtheorem{rqc}{RQ-C}
\newtheorem{rqg}{RQ-G}
\newtheorem{rqbv}{RQ-BV}

\usepackage{adjustbox}
\usepackage{tikz}
\usetikzlibrary{calc}
\usetikzlibrary{matrix}
\usetikzlibrary{positioning}
\usetikzlibrary{shapes.geometric}
\usetikzlibrary{shapes.misc}
\usetikzlibrary{arrows}
\usetikzlibrary{calc}
\usetikzlibrary{shapes.multipart}
\usetikzlibrary{decorations.pathreplacing,arrows.meta,calligraphy}
\usetikzlibrary{decorations.text}
\begin{document}
\title{Process Analytics -- Data-driven Business Process Management}

\author[1]{\fnm{Matthias} \sur{Stierle}}\email{matthias.stierle@fau.de}
\author[2]{\fnm{Karsten} \sur{Kraume}}
\author[1]{\fnm{Martin} \sur{Matzner}}

\affil[1]{\orgdiv{Chair of Digital Industrial Service Systems}, \orgname{Friedrich-Alexander Universit\"at Erlangen-N\"urnberg}, \country{Germany}}
\affil[2]{\orgdiv{ERCIS}, \orgname{University of M\"unster}, \country{Germany}}

\abstract{Data-driven analysis of business processes has a long tradition in research. However, recently the term of process mining is mostly used when referring to data-driven process analysis. As a consequence, awareness for the many facets of process analysis is decreasing. In particular, while an increasing focus is put onto technical aspects of the analysis, human and organisational concerns remain under the radar. Following the socio-technical perspective of information systems research, we propose a new perspective onto data-driven process analysis that combines the process of analysis with the organisation and its stakeholders. This paper conceptualises the term process analytics and its various dimensions by following both an inductive and deductive approach. The results are discussed by contrasting them to a real-life case study from a large company implementing data-driven process analysis and automation.}

\maketitle    

\section{Introduction}
\label{sec:intro}

Analytics, defined as a methodology to create business value from data, has been shown to have a positive effect on process performance \citep{TRKMAN2010318}. As such, process data and its analysis as well as further applications that are built on top of it, are crucial to optimise the performance of business processes \citep{davenport2018from}.
Therefore, it comes as little surprise that process analysis and in particular data-driven approaches have a long history in \ac{BPM} research. Many research streams exist that are dedicated to technologies making use of process data. \Ac{PW} for instance deals with the question of how to store and query process data in a structured way \citep{grigori2004business}. In contrast, \ac{BAM} is concerned with the analysis of real-time event streams for process monitoring \citep{janiesch2012beyond}. A more recent stream is \ac{PBPM} which builds on historical data to predict future process behaviour \citep{marquez2017predictive}.
Despite the plurality of research streams, almost all of the recent research efforts dealing with the use of process data are categorised as \emph{process mining}. Process mining is an evidence-based technology for the discovery, conformance checking, and enhancement of business processes based on process data (so-called event logs) \citep{VanDerAalst2011}. Over the last decade, a fast-growing number of publications have used the term \emph{process mining}\footnote{see Scopus results on \emph{TITLE-ABS-KEY ( "process mining" )}}, and similarly, commercial vendors were not shy to claim \emph{process mining} capabilities \citep{viner2703process}. Process mining seems to have become a universal term for the analysis of process data --- far away from the initially defined types of process mining \emph{discovery, conformance checking and enhancement} concerned with the analysis of historical process data \citep{VanDerAalst2011}.
Consequently, with this fast-paced development, it became somewhat unclear what process mining is and what it is not. And in particular, how process mining fits into the larger picture of process analysis which is broader as it does not limit itself to the technical approaches.

The focus on process mining has possibly led to a focus on the technological aspects of process analysis. As pointed out by \citet{VanDerAalst2016}, the research community has sometimes lost track of the ultimate goal of process analysis, and \ac{BPM} respectively, which is improving the process. Instead, the focus was put, for instance, on creating semantically correct process models from data \citep{VanDerAalst2016}. A large share of research has presented techniques and methods to process event logs but has neglected further aspects of analytics with \textquote{IS flavour}\citep{power2018defining}, such as governance or culture \citep{cosic2015business}, which are crucial for successful application in organisations. A similar observation is made by \citet{Hassan2020where} for analytics in general. He concludes that much effort is spent on creating and redefining algorithms, but too little on studying the impact of their deployment. Simultaneously, the results of the recent 
\emph{SIM IT Trends Study} show that making analytics a success in organisations is still among the top three concerns of managers \citep{kappelman2020trends}.

This observation calls for a new perspective that considers both technical and socio-technical aspects for improving business processes based on process data. The term of (business) process analytics has been used various times by scholars \citep{schwegmann2013method,zur2015business,polyvyanyy2017process,mehdiyev2020prescriptive}, but not deliberately as it seems. The focus is mostly put on technology and the process of analysis \citep[\textquote{approaches, techniques and tools},][]{polyvyanyy2017process}), but human and organisational concerns are widely ignored. Thus, a sound definition of process analytics in line with established definitions of (business) analytics is missing, and so is a broader conceptualisation of the topic. Various scholars point out the importance of defining concepts as a basis for developing theory and artefacts \citep{frank2006towards} in IS research. Further, \citet{Hassan2020where} calls for analytics research to build such theory to better understand the impact of data.

Accordingly, the primary goal of this work is to derive a well-grounded definition of the term \emph{process analytics} and to present essential constructs of the domain \citep{march1995design}. The aim is to raise awareness of the importance of the various dimensions of process analytics besides technology, which will guide scholars in future research efforts and provide practitioners with a list of aspects to consider when implementing process analytics. Moreover, this research aims to point out the variety of available technologies for process analysis.

To achieve the research objective, both an inductive and deductive approach are chosen. Existing works on (business) analytics are reviewed, and the identified concepts are then transferred to the \ac{BPM} domain. Furthermore, to assure the practical relevance, insights about process analytics are gathered from semi-structured expert interviews and publicly available case studies. The results gained from the knowledge base and the cases are used for defining the term process analytics and its constructs. These results are discussed and demonstrated with a case study from an industry partner.

The remainder of this paper is organised as follows. Section \ref{sec:background} presents the results of a literature review on analytics and the preliminaries of \ac{BPM}. In Section \ref{sec:interviews}, the method, results and a summary of the conducted interviews and the reviewed case studies are presented. In Section \ref{sec:pa}, the term of process analytics is defined, and relevant dimensions are discussed. Section \ref{sec:case-arvato} reports a case study from implementing evidence-based process analysis and automation at Arvato CRM solutions to demonstrate the utility of a conceptualisation of process analytics and its various dimensions. Section \ref{sec:discussion} discusses contributions, limitation and implications for future research. Last, Section \ref{sec:conclusion} concludes the paper with a brief summary.



\section{Background}
\label{sec:background}

\subsection{Understanding Analytics}

Two perspectives onto the term \emph{analytics} present themselves: the etymological meaning and the phenomena that have occurred which correspond to today's meaning of \emph{analytics}.
The ancient origin of analytics stems from the Greek word \emph{analȳtikós} \citep{power2018defining,hassan2019origins} and, similar to the word \emph{analysis}, means \textquote{unloosen}\footnote{https://www.etymonline.com/word/analytics}. The etymological origin suggests \emph{analytics} and \emph{analysis} have the same meaning, which, however, is not the case \citep{delen2018research}. 
Analysis has been defined as \textquote{resolution of anything complex into simple elements}\footnote{https://www.etymonline.com/word/analysis} in the 16\textsuperscript{th} century to provide a better understanding of a matter. More recent definitions describe it as \textquote{the process of studying or examining something in an organised way to learn more about it}\footnote{https://dictionary.cambridge.org/dictionary/english/analysis}. In contrast, the suffix \textquote{-ics} indicates that a word describes a body of knowledge and principles \citep{power2018defining}. So while analytics includes analysis, or rather methods and techniques for analysis, it covers further aspects and has been considered a \textquote{methodology that encompasses a multitude of methods and practices} \citep{delen2018research}.

According to \citet{delen2018analytics}, the term (business) analytics has only been used on a broad scale since the 2000s. Before, terms such as \emph{decision support systems}, \emph{business intelligence} or, most recently, \emph{big data} have been used to describe what is understood as analytics: 
\emph{employing internal or external, structured or unstructured data for actionable insights} \citep{delen2018analytics}. \citet{holsapple2014unified} follow a similar line of argumentation and describe business analytics as \textquote{operating on data, with an aim of supporting business activities (e.g., decision making)}. The latter definition shows that today's understanding of analytics goes beyond decision support, and data can be used to create value in various ways. 
\citet{davenport2018from} describes how analytics has evolved from internal decision support (business intelligence) over \textquote{data products} to analytics-driven business models and is yet to further evolve with the use of \ac{AI} for automation (Analytics 4.0).
According to \citet{holsapple2014unified}, reasons, why organisations engage in analytics, are amongst others creating a competitive advantage, improving organisational performance, better decisions or generally speaking obtaining value from data.

To further structure the concept of analytics, scholars have presented various dimensions of analytics. While there is no exact consensus about the integral dimensions of conceptualising analytics, many works share a common view, as shown in Table \ref{tab:concepts_analytics}.

\begin{table}																																					
\caption{Matrix of Business Analytics Concepts.}																																		
\label{tab:concepts_analytics}		
\centering\settowidth\rotheadsize{Business Value/}																																					
\renewcommand\cellalign{cc}																																					
\renewcommand\arraystretch{1.25}																																					
\begin{tabular}{|p{5cm}|>{\centering\arraybackslash}p{0.5cm}|>{\centering\arraybackslash}p{0.5cm}|>{\centering\arraybackslash}p{0.5cm}|>{\centering\arraybackslash}p{0.5cm}|>{\centering\arraybackslash}p{0.5cm}|>{\centering\arraybackslash}p{0.5cm}|>{\centering\arraybackslash}p{0.5cm}|>{\centering\arraybackslash}p{0.5cm}|}

\hline																																					
																																					
\diagbox[height=1.25\rotheadsize, innerwidth=5cm]{\textbf{Articles}}{\textbf{Concept}}																																					
	 & \rotcell{Technology}	 & \rotcell{Data}	 & \rotcell{Business Value}	 & \rotcell{Governance}	 & \rotcell{Culture}	 & \rotcell{People}																															
\\ \hline																																					
																																					
\citet{Wixom2013MaximizingVF}	 & \textbullet	 & \textbullet	 & \textbullet	 & \textbullet	 &	 &	\\ \hline																														
																																					
\citet{banerjee2013data}	 & \textbullet	 & \textbullet	 &	 &	 & \textbullet	 &	\\ \hline

\citet{holsapple2014unified}	 & \textbullet	 &	 & \textbullet	 & \textbullet	 & \textbullet	 &	\\ \hline

\citet{cosic2015business}	 & \textbullet	 &	 &	 & \textbullet	 & \textbullet	 & \textbullet	\\ \hline

\citet{Seddon2012HowDB}	 & \textbullet	 & \textbullet &	 & \textbullet	 &	 & \textbullet	\\ \hline																														
\citet{janssen2017factors}	 &	 & \textbullet	 &	 & \textbullet	 &	 & \textbullet	\\ \hline

\citet{power2018defining}	 & \textbullet	 &	 & \textbullet	 & \textbullet	 &	 &	\\ \hline

\citet{davenport2018from}	 & \textbullet	 & \textbullet	 &	 &	 & \textbullet	 & \textbullet	\\ \hline

\citet{delen2018research}	 & \textbullet	 & \textbullet	 & \textbullet	 &	 & \textbullet	 & \textbullet	\\ \hline

\citet{hassan2019origins}	 &	 & \textbullet	 & \textbullet	 & \textbullet	 &	 &	\\ \hline																														
\citet{oneill2019business}	 & \textbullet	 &	 & \textbullet	 & \textbullet	 & \textbullet	 & \textbullet	\\ \hline

																											\citet{vandewetering2019big}	 & \textbullet	 &	\textbullet &  & \textbullet	 & \textbullet	 & \textbullet	\\ \hline

\end{tabular}																																					

\end{table}

\begin{description}
\item[Technology] 

As expected, the most commonly mentioned dimension is techniques, tools and algorithms for analysis. Technology comprises the technical capabilities of organisations to execute analytics. A popular taxonomy of analytics' techniques is the differentiation between descriptive, diagnostic, predictive or prescriptive analytics \citep{banerjee2013data,delen2018analytics}. Similarly, \citet{davenport2010analytics} present key questions of analytics and further separate them in providing (known) information or (new) insights (see Table \ref{tab:keyquestions}). Descriptive analytics answers questions such as \emph{What happened?} and \emph{What is happening now?}. Diagnostic analytics seeks answers to \emph{How and why did it happen?}. Predictive analytics is concerned with \emph{What will happen?} while prescriptive analytics aims at finding out \emph{What is the next best action?} and \emph{What is the best/worst that can happen?}.

\begin{table}[ht]
\centering
\caption{Key questions addressed by analytics \citep{davenport2010analytics}}
\label{tab:keyquestions}
\begin{tabular}{|r|C{3.5cm}|C{3.5cm}|C{3.5cm}|}
\hline
 & \textbf{Past} & \textbf{Present} & \textbf{Future} \\ \hline
\textbf{Information} & What happened? & What is happening now? & What will happen? \\ \hline
\textbf{Insight} & How and why did it happen? & What is the next best action? & What is the best/worst that can happen? \\ \hline
\end{tabular}

\end{table}

\item[Data] 
Surprisingly, data (and its management) is often not mentioned explicitly despite its inherently essential roll for analytics. Aspects that are discussed are the type of data \citep[e.g. structured or unstructured,][]{hassan2019origins}, data architectures and integration, data quality as a key for analytics \citep{banerjee2013data} success as well as data security and privacy concerns \citep{janssen2017factors}.
\item[Business Value/Objectives] 
Various scholars describe defining (and measuring) business value of an analytics effort as crucial for success \citep{power2018defining}. Furthermore, many mention the importance of aligning the objectives with the overall objectives of the organisation. 

\item[Governance] 
A frequently mentioned aspect of analytics is governance, together with processes for the implementation of techniques and the transformation of the organisation. 
Governance is closely linked to many other dimensions such as monitoring and aligning the objectives \citep{cosic2015business}, sharing best practices about technologies or compliant data access.
\item[Culture]
Establishing an analytics culture that fosters evidence-based decision-making instead of \emph{gut feeling} is a key challenge for organisations. \citet{holsapple2014unified} call for the need of a \emph{movement} to change mindsets while \citet{cosic2015business} see analytics culture as \textquote{organisational norms, values and behavioural patterns that form over time}.
\item[People] 
The people dimension of analytics is manifold as it covers all individuals involved \citep{cosic2015business}. On the one hand, it comprises the users which range from managers with analytics thinking over business analysts up to operational users being supported by analytics systems. On the other hand, many further stakeholders might be involved, such as IT experts or customers.  Recruiting is an essential part of the analytics journey, but just as much attention needs to be put into training the workforce and change management.
\Citet{janssen2017factors} find that collaboration and knowledge exchange between various stakeholders and departments in the organisation is a key success factor for analytics. 
\end{description}

\subsection{Analytical Applications in Business Process Management}

A business process is defined as a \textquote{timely and logical sequence of activities} \citep[p.4]{becker2012}.
More generally speaking, processes are essential for organisations to deliver a service or a product to a customer \citep{dumas2013fundamentals}. The efficient design of high-quality processes has only received major awareness since the 1980s \citep{dumas2013fundamentals}. \Ac{BPR} became a generally accepted task for managers which later on developed into \ac{BPM} -- \textquote{a well-established set of principles, methods and tools that combine knowledge from information technology, management sciences and industrial engineering with the purpose of improving business processes}\citep{VanDerAalst2016}.

A well-accepted conceptualisation of \ac{BPM} activities are life-cycle models \citep{dumas2013fundamentals,hammer2015business,vanderAalst2016book}, which --- while being different --- mostly share these common phases: process analysis/diagnosis, process (re-)design, process implementation and process monitoring/controlling. 

\begin{figure}
    \centering
    \includegraphics[width=.8\textwidth]{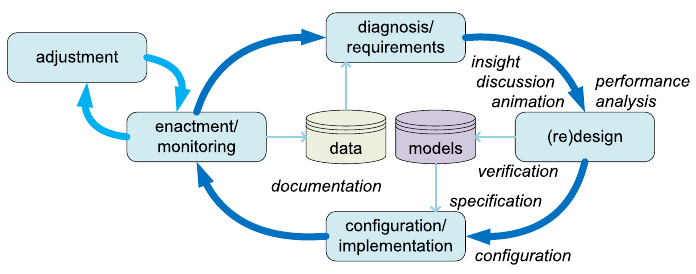}
    \caption{BPM life-cycle model by \citet[p.31]{vanderAalst2016book}}
    \label{fig:bpmlc_vda}
\end{figure}

As such, data as a basis for analysis and monitoring has long been a crucial part of \ac{BPM} as presented in the life-cycle model in Figure \ref{fig:bpmlc_vda}.

In particular, process mining has found major application both in research and in practice \citep{davenport_2020}. Process mining is a technology -- that is, a set of techniques and methods -- for the analysis of business processes using event data retrieved from \acp{IS} \citep{van2012process}. \Citet{van2012process} distinguishes mainly between three types of process mining. 
Process discovery aims at generating process models from event logs. Conformance checking is used for identifying deviations between defined to-be process models and actual process behaviour (derived from an event log). Last, process enhancement is dedicated to improving existing process models based on observed, past behaviour.


However, many other research streams concerned with the analysis and use of process data exist. For instance, one path of research has been dedicated to \ac{RCA} for process performance deviations \citep{suriadi2012root,Lehto2016} which has most recently been complemented by causal inference techniques \citep{Narendra2019}. Also, \ac{BAM} deals with real-time analysis of process executions for process monitoring and controlling \citep{janiesch2012beyond}. In recent years, \ac{PBPM} has been established as a combination of \ac{ML} and process monitoring techniques \citep{marquez2017predictive} to foresee future process behaviour. Various scholars have proposed to go beyond predictive to prescriptive process control by suggesting counteractions to the process workers \citep{krumeich2016prescriptive,mehdiyev2020prescriptive}. Finally, \ac{RPM} is a novel domain for identifying process automation opportunities from user interaction data \citep{leno2020robotic}.

What unites all these research streams is their focus on the process of analysis --- that is, transforming process data into information and insights --- which used to be summarised as \textquote{business process intelligence} \citep{grigori2004business}.

Little attention, however, has been paid to achieving a value gain with this information in an organisational setting, i.e. considering the socio-technical perspective. Only very recently, scholars in the field of \ac{BPM} have been raising questions about value realisation, adoption and use of technologies for process analysis and monitoring --- in particular, in the context of process mining \citep{eggers2020turning,grisold2020adoption,mci/Brocke2021} --- but the answers have yet to be found.


\section{Interviews}
\label{sec:interviews}
\subsection{Method and Data}
To better understand what aspects are relevant for practitioners and organisations besides technology to apply process analytics successfully, and to match the insights gained from the knowledge base on business analytics, a qualitative approach was chosen. Semi-structured interviews were conducted with six people that were experienced in executing process analytics, i.e. applying process mining in an organisation (see Table \ref{tab:interviewees}). Given the diversity of process stakeholders in organisations, we chose interviewees that represented various roles such as end users, IT and process owners. While an interview guide (adjusted per role) was used to structure the interviews (see Appendix \ref{appendix:sec:interview-guide}), a large degree of freedom was given to the interviewees to report on their process mining \emph{journey} and to talk about aspects that they considered necessary.

\begin{table}[h]
\centering
\caption{Overview of Interviewees}
\label{tab:interviewees}
\begin{tabular}{|l|l|c|l|}
\hline
\textbf{Role} & \textbf{Industry} & \textbf{\# employees} & \textbf{Source} \\ \hline
Process Owner & Computer software & \textgreater 10.000 & Interview \\ \hline
Topic Owner & Manufacturing & \textgreater 10.000 & Interview \\ \hline
End User & Engineering & \textgreater 10.000 & Interview \\ \hline
Process Owner & Telecom & 1.000 - 10.000 & Interview \\ \hline
Consultant & IT Consulting & \textless 100 & Interview \\ \hline
IT & Telecom & \textgreater 10.000 & Interview \\ \hline

\end{tabular}
\end{table}

Further, the results were extended by reviewing eight case studies that were available online. Namely, the case studies were presented at \emph{Celosphere}\footnote{https://www.celonis.com/celosphere-live} --- a commercial process mining event --- and/or in the book \emph{Process Mining in Action} \citep{reinkemeyer2020process} (see Table \ref{tab:cases}).
Summing up, 14 cases were collected from people in various roles working for different organisations in different industries globally.
All of the interviews and case studies were video and audio recorded, transcribed, and further analysed using the open-coding technique assigning summative words or short, essence-capturing phrases to the text-based data to highlight the most prominent themes within several interviews \citep{saldana2015coding}. 

\begin{table}[h]
\centering
\caption{Overview of Case Studies}
\label{tab:cases}
\begin{tabular}{|l|l|l|c|p{3.5cm}|}
\hline
\textbf{Role} & \textbf{Company} & \textbf{Industry} & \textbf{\# employees} & \textbf{Source} \\ \hline
		Top Mgmt &  Reckitt Benckiser & Consumer goods &  	$\approx$ 40.000 & Celosphere \\ \hline
		Top Mgmt & Deutsche Telekom & Telecom & 	$\approx$ 200.000 & Celosphere \\ \hline
		Topic Owner & Lufthansa CityLine & Airline & 	$\approx$ 2.000 & Celosphere \\ \hline
		IT &  Plansee SE & Manufacturing & 	$\approx$ 14.000 & Celosphere \\ \hline
		IT & Siemens & Electronics & 	$\approx$ 385.000 & Celosphere + \citet{reinkemeyer2020process} \\ \hline
		Topic Owner & Avnet & Electronics & 	$\approx$ 15.000  & Celosphere \\ \hline
		IT & ABB & Electronics & 	$\approx$ 150.000 & Celosphere + \citet{reinkemeyer2020process}  \\ \hline
		IT & VTB Bank &  Banking & 	$\approx$ 80.000 & Celosphere \\ \hline
\end{tabular}
\end{table}

\subsection{Results}

Overall the insights that we retrieved from the cases can be categorised into triggers and (expected) outcomes (i.e. benefits) of process mining implementation, achieved benefits as well as challenges and success factors. In the following, we will focus on the latter and present the most mentioned success factors.

The most commonly mentioned aspect for the successful application of process mining technology in an organisation was highlighting the technology's value, i.e. its expected benefit (9)\footnote{The value 9 indicates that this aspect was mentioned nine times in the reviewed cases and the conducted interviews.}. Many cases reported the missing vision about process analytics opportunities during the setup phase as problematic (8). Pointing out the objectives and the potential value of the technology was both mentioned as necessary to give the effort meaning and direction, and foster acceptance and adoption by users. Several cases reported on the importance of motivating users (8) together with enabling them (7), including offering training and educational sources (6). Clear governance structures (2), together with collecting best practices (4), is considered relevant. One interviewee and two case studies suggested establishing a \emph{centre of excellence} to provide guidance with methods and tools. Some cases mentioned data quality (4) as a potential issue, but more importantly, the significance of data integration and access was stressed (8). In that regard, it was also commonly mentioned to involve IT experts (9). In general, enforcing collaboration between departments (5) and, in particular, involving those people with process knowledge (8) was considered to be a crucial factor for successfully applying process mining. One of the interviewees revealed that to decide whom to invite for a workshop, they send out a \textquote{puzzle} to potential participants to evaluate their process thinking skills and process knowledge. Another interviewee pointed out that in hindsight, he would engage process experts from various departments in the very beginning.
Last, in several cases, a key benefit achieved with process mining was to move the discussion from subjective opinions to objective facts (6). Several interviewees, however, pointed out that establishing evidence-based decision making as a paradigm in the organisation was cumbersome and required a change in company culture. 

    


\subsection{Conclusion}
The insights gained from the cases largely confirm the results collected from the knowledge base about analytics. However, some aspects were mentioned which are interesting specifically for process analytics.
First of all, neither in the interviews nor in the case studies aspects regarding the technology dimension were mentioned frequently. A simple reason might be that process mining techniques, methods and algorithms are embedded into commercial software. In contrast to, for instance, data mining where users need to build, train and deploy models, the models or rather algorithms in process mining are not accessible to users. While this insight might be worthwhile for further analysis, a certain bias must be admitted as in almost all cases the same commercial process mining software was used.

A more exciting and substantial finding was the importance of collaboration between different departments and stakeholders with process knowledge including IT, data scientists, process workers and managers/process owners. Process mining often assumes the existence of process data and a reference process model (e.g. for conformance checking) that documents process knowledge. Not all information about a process, however, is contained in such models (reasons might be that the model is outdated, the modelling notation does not support the type of information, or simply that no model exists). Therefore, gathering such process knowledge to execute a process analysis or correctly interpret its results seems to be of similar importance as the (event) data itself.







\section{Defining Process Analytics}
\label{sec:pa}

From what we have learned, process analytics can be defined as the body of knowledge, principles, techniques and methods to obtain value from business process data. In particular, process analytics combines the process of the analysis (e.g. using process mining) with human and organisational concerns which makes it a well-fitting topic for \ac{IS} research \citep{hassan2019origins}. 

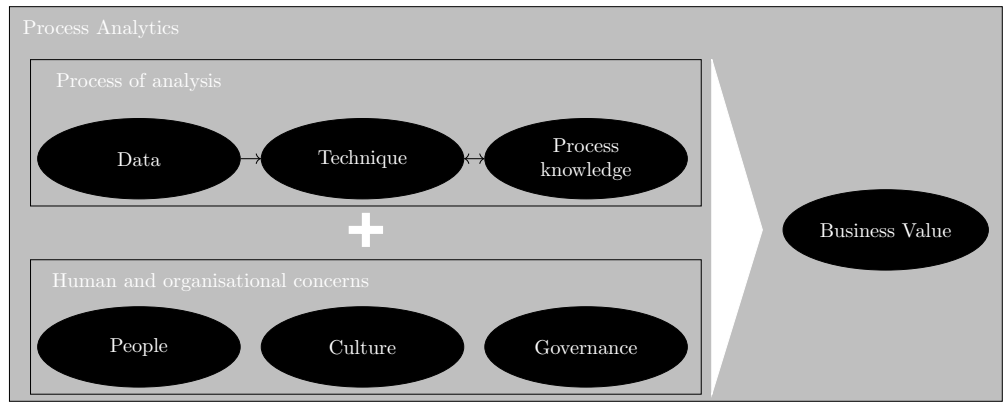
\begin{figure}[h]
    \centering
    \tikzstyle{box} = [draw, fill=lightgray, rectangle, 
    minimum height=4em, minimum width=8em, align=center]
\tikzstyle{boxtext} = [draw, fill=black,text=white, ellipse, 
    minimum height=4em, minimum width=10em, align=center]
\tikzstyle{labelnode} = [  text=white, align=left, minimum height=1em]
\tikzset{cross/.style={cross out, draw=white, minimum size=2*(#1-\pgflinewidth), inner sep=0pt, outer sep=0pt},
cross/.default={1pt}}
\begin{adjustbox}{width=\textwidth}
\begin{tikzpicture} %
[align=center,node distance=1em,every label/.append style={font=\footnotesize}]]
    \node (pa) [box] {
        \begin{tikzpicture}
        [align=center,node distance=1em,every label/.append style={font=\footnotesize}]]
            \node (poa) [box] {
                    \begin{tikzpicture}
                        \node (labelpoa) [labelnode,anchor=south east] at (0,0) {Process of analysis};
                        \node (data) [boxtext,below=of labelpoa] {Data};
                        \node (teq) [boxtext, right=of data] {Technique};
                        \node (output) [boxtext, right=of teq] {Process \\ knowledge};
                        \draw [->] (data) -- (teq);
                        \draw [<->] (teq) -- (output);
                    \end{tikzpicture}
                    };

            \node (hoc) [box,below=6em of poa,anchor=center] {
                    \begin{tikzpicture}
                        \node (labelhoc) [labelnode,anchor=south east] at (0,0) {Human and organisational concerns};
                        \node (people) [boxtext,below=of labelhoc,anchor=north east] {People};
                        \node (culture) [boxtext, right=of people] {Culture};
                        \node (gov) [boxtext, right=of culture] {Governance};
                    \end{tikzpicture}
                    };
            \node (plus) [cross=10pt,anchor=center, rotate=45,line width=4pt] at ($(poa)!0.5!(hoc)$) {};
            \node (value) [boxtext,xshift=2em, anchor=west] at ([xshift=2em]poa.east  |- plus){Business Value};
            \node (labelpa) [labelnode,above left=of poa,anchor=south west] {Process Analytics};
            \draw [transform canvas={xshift=.5em},white, fill=white] (poa.north east) -- (hoc.south east)-- ([xshift=2.5em]poa.east  |- plus) -- (poa.north east);
        \end{tikzpicture}
    };
\end{tikzpicture}
\end{adjustbox}
    \caption{Process Analytics and relevant dimensions}
    \label{fig:pa_concepts}
\end{figure}

Technologies such as process mining, \ac{BAM} or \ac{PBPM} are essential for transforming process data into information and, subsequently, knowledge.
However, both the knowledge base on analytics and the reviewed cases indicate that organisations need to consider various dimensions to achieve a value gain. \ac{BPM} research should hence investigate the various facets of making process analytics successful. In the following, we will outline the dimensions that we consider most important for process analytics, including some exemplary research questions, following the socio-technical perspective, as suggested by \citet{bostrom1977mis}. This conceptualisation of process analytics depicted in Figure \ref{fig:pa_concepts} shall serve as a starting point for scholars and stimulate a discussion about process analytics in BPM research.

As pointed out, much research has been dedicated to the \emph{process of analysis} -- that is, transforming process data into information and insights. In particular, a multitude of techniques and methods based on process data have been designed by scholars. As shown in Table \ref{tab:keyquestions_pa}, the portfolio of technologies ranges from descriptive process analytics over diagnostic analytics up to predictive and prescriptive process analytics. Without a doubt, the design of new techniques and methods for process analytics will remain a crucial task for BPM research.
\begin{table}[]
\centering
\caption{Key questions addressed by process analytics technologies based on \citet{davenport2010analytics}}
\label{tab:keyquestions_pa}
\begin{tabular}{r|p{3.5cm}|p{3.5cm}|p{3.5cm}|}
\textbf{} & \textbf{\textbf{Past}} & \textbf{\textbf{Present}} & \textbf{\textbf{Future}} \\ \hline
\multicolumn{1}{|r|}{\multirow{2}{*}{\textbf{Information}}} & \footnotesize{What happened?} & \footnotesize{What is happening now?} & \footnotesize{What will happen?} \\
\multicolumn{1}{|r|}{} & Process Mining 
& Business Activity Monitoring & Predictive Process Monitoring \\ \hline
\multicolumn{1}{|r|}{\multirow{2}{*}{\textbf{Insight}}} & \footnotesize{How and why did it happen?} & \footnotesize{What's the next best action?} & \footnotesize{What's the best/worst that can happen?} \\
\multicolumn{1}{|r|}{} & Root Cause Analysis & Prescriptive Process Monitoring & Causal Inference \\ \hline
\end{tabular}
\end{table}
Moreover, organisations usually introduce technology with software products or tools. Hence, process analytics software evaluation, selection, and implementation should be a concern \citep{viner2703process}. For instance, criteria need to be defined to compare software, and reference processes are required for the selection.

\subsubsection*{Exemplary research questions concerning technology}

\begin{rqt}\label{rqt1}How to implement techniques for prescriptive process analytics?\end{rqt}
\begin{rqt}\label{rqt2}What are the criteria to evaluate and select process analytics software for use in organisations?\end{rqt}
\begin{rqt}\label{rqt3}How to develop, test and deploy process analytics techniques in organisations?\end{rqt}

While the literature has also discussed the \emph{data} perspective (e.g. XES standard for event logs), most work has assumed the existence of process data in the form of (clean) event logs as a starting point. A recent trend towards object-centric process mining \citep{van2019object} acknowledges that for organisations, process analysis starts with information systems and underlying databases instead of readily available event logs. Furthermore, \ac{RPM} is among the first to discuss explicit logging and tracking of process data instead of assuming the existence of extensive event logs \citep{leno2020robotic}. Last, data privacy concerns should be discussed both on a technical level \citep{rafiei2020privacy} and through the organisational lens. 

\subsubsection*{Exemplary research questions concerning data}

\begin{rqd}\label{rqd1}How to integrate and analyse process data that spans various systems and processes?\end{rqd}
\begin{rqd}\label{rqd2}How to log data for process analytics without introducing a bias?\end{rqd}
\begin{rqd}\label{rqd3}What can be achieved by technical means to assure data privacy in process analytics and which organisational policies and procedures could be useful?\end{rqd}

The purpose of process analysis is to derive \emph{process knowledge}, and at the same time, it can depend on process knowledge as input (e.g. conformance checking). To give meaning to the insights gained from the data, knowledge about the context of the business process is crucial, which is often not contained in the data \citep{rosemann2008contextualisation}. As the most prominent method for data mining, the CRISP-DM method suggests business/domain understanding to be the first step \citep{shearer2000crisp}. A recent trend in \ac{ML} research is to combine user input and \ac{ML} models \citep{pearl2019seven}. Hence, it seems plausible that process analysis should begin with an understanding of the process, and subsequently, process knowledge plays an integral part in process analytics. Questions around how to gather, preserve, share and most importantly act on the derived knowledge arise. 

\subsubsection*{Exemplary research questions concerning process knowledge}

\begin{rqpk}\label{rqpk1}How to convert information and insights gained through process analysis into knowledge and actions?\end{rqpk}
\begin{rqpk}\label{rqpk2}How to record and share derived knowledge in the organisation?\end{rqpk}
\begin{rqpk}\label{rqpk3}How to gather knowledge about a business process and its context required for initiating a process analytics project?\end{rqpk}

Besides the \emph{process of analysis}, process analytics needs to consider \emph{human and organisation concerns} to achieve value. This socio-technical perspective is the core of IS research and, hence, it has many facets. To maintain a manageable level of complexity, we will limit this view to three aspects: the individuals in the organisation, i.e. the \emph{people}; the organisation's \emph{culture}, including leadership as a manifestation of how the individuals work together; \emph{governance} as the processes which control how individuals work together.

A key aspect of business processes is that they extend across the organisation and a multitude of \emph{people} are typically involved in \ac{BPM} activities such as managers, process owners, process participants, process analysts, system engineers or process consultants \citep{dumas2013fundamentals}, and, ultimately, also customers. Understanding which roles are required for process analytics and how these can contribute to a successful project is crucial. Many challenges arise such as acceptance, adoption, usage and integration of technologies into users' work routines. Understanding the needs and expectations of various roles can guide the design of techniques for process analysis \citep{seeliger2019processexplorer,stierle2020design}.

\subsubsection*{Exemplary research questions concerning people}

\begin{rqp}\label{rqp1}Which roles and skills are required for process analytics? How do job profiles change?\end{rqp}
\begin{rqp}\label{rqp2}How to integrate process analytics in employee' routines?\end{rqp}
\begin{rqp}\label{rqp3}How to design process analytics techniques for increased adoption and sustainable usage?\end{rqp}


Certainly, process mining can provide valuable information and insights. However, acting on these requires a company \emph{culture} that embraces evidence-based decision making. Experienced process workers might tend to rely on \emph{gut feeling} rather than using the information provided by a process analytics system. Also, existing process models and rules might be preferred for decision-making. However, digital transformation entails \textquote{light touch processes} \citep{abayomi2020digital} which require more flexibility in decision-making. Bringing together process compliance, user experience, and process analytics for decision-making in processes could be a major challenge for organisations. Furthermore, process data contains sensitive information about employees and their behaviour, which is made transparent by technologies such as process mining. A culture that encourages the use of process analytics needs to emphasise the improvement of the organisation and its business processes \citep[i.e. process innovation,][]{mikalef2020examining} instead of punishing mistakes on an individual level \citep{hammer2015business}. Hence, the existing organisational culture has a big influence on process analytics success and needs to be understood. Summing up, a consensus is required on how to use process data in the organisation and how to act on the insights gained. 

\subsubsection*{Exemplary research questions concerning culture}

\begin{rqc}\label{rqc1}How to establish an organisational culture that fosters evidence-based decision making in business processes?\end{rqc}
\begin{rqc}\label{rqc2}How to deal with an existing organisational culture that hinders adoption of process analytics?\end{rqc}
\begin{rqc}\label{rqc3}How to create trust in an organisation's handling of sensitive process data?\end{rqc}
\begin{rqc}\label{rqc4}Which norms and values are important for a culture that fosters process analytics?\end{rqc}

In addition, \emph{governance} is a concern --- mostly of practical nature --- when implementing process analytics. New governance structures need to be implemented or existing governance structures need to be adjusted to steer and monitor the activities \citep{grisold2020adoption}. For example, the concept of a centre of excellence has been discussed before in \ac{BPM} research \citep{rosemann2015service} and has also found success in process mining implementations \citep[p.172]{reinkemeyer2020process}. In contrast to culture, which is based on values and norms, governance needs to provide processes and tools for individuals working together. 

\subsubsection*{Exemplary research questions concerning governance}

\begin{rqg}\label{rqg1}Which organisational structures are useful for process analytics?\end{rqg}
\begin{rqg}\label{rqg2}How to incorporate process analytics in existing organisational structures and routines?\end{rqg}
\begin{rqg}\label{rqg3}How to design reference processes for process analytics implementation?\end{rqg}

Finally, the key concern of process analytics should be on its purpose, which is creating \emph{business value}. \citet{eggers2020turning} stress the relevance of understanding value realisation from process mining usage and present a research agenda to investigate the value creation process before, during and after implementation. \citet{grisold2020adoption} highlight the importance of measuring and assessing efforts vs costs for process mining initiatives. 
The question of purpose and value should also guide the design of process mining techniques. For instance, many works in the \ac{PBPM} domain neglect the intended application and design generic techniques. \citet{mehdiyev2020explainable} show how defining expected outcomes of a technique facilitate design decisions when creating process mining artefacts. Most works in the domain follow the design science research paradigm, but the usefulness of the created artefacts for organisations is rarely evaluated.

\subsubsection*{Exemplary research questions concerning business value}

\begin{rqbv}\label{rqbv1}Which factors do influence value creation through process analytics?\end{rqbv}
\begin{rqbv}\label{rqbv2}How to measure costs and benefits of process analytics?\end{rqbv}
\begin{rqbv}\label{rqbv3}How to design useful techniques that contribute to value creation?\end{rqbv}





\section{Process analytics at Arvato CRM solutions}
\label{sec:case-arvato}

\subsection{Motivation for the case study}
Having conceptualised process analytics and its relevant dimensions, we will now discuss the framework in the context of Arvato CRM solutions (subsequently referred to as "the company") and go through different dimensions in a case study style. Here we focus on dimensions that are often neglected in the context of process analytics, yet mission critical from our experience both as practitioners and scholars. Utilisation of technologies (such as process mining) in an organisation depends on people and subsequently requires governance. While a fast implementation might be very appealing to consultants and engineers that strive for short-term optimisation, it is important to consider that process analytics alters the work environment of many agents \citep{tang2020creating}.  Therefore, process analytics has to be positioned adequately through communication and change management to avoid negative side effects.

The company lends itself as a case study for three reasons. First, being a business process outsourcing (BPO) provider, successful application of process analytics is crucial to compete in the market. Second, the company has changed (organisational) governance multiple times coming from conglomerate, transforming into matrix and ultimately becoming a line organisation. Thus, implementing new technologies (for process analysis and automation) is a challenge. Third, the company is one of the most international entities of Bertelsmann SE and therefore has to handle cultural complexities not only given the change in organisational structure but also given business operations across countries and corresponding language barriers and differing approaches to business, process management and utilisation of data.   Thus, the company has complex organisational structures that create substantial challenges when defining analytics-related governance on top of other structures and procedures. Thus, the successful adoption of process analytics --- which is of high strategic importance --- is at risk.

\subsection{Introduction of company and its structure}
As of today (2021) the company is legally not existing any more. It has been carved out from Arvato AG which till date is part of Bertelsmann SE. Since 2020 the company has operated under the corporate brand Majorel and has recently been listed at
EURONEXT\footnote{https://www.euronext.com/en/about/media/euronext-press-releases/majorel-lists-euronext-amsterdam}.
When we elaborate on the company we focus on the period between 2016 and 2018 where it provided customer services on behalf of its business to business (B2B) clients from different industries ranging from telecommunication, banking, insurance, over automotive and healthcare to public sector. 

From a functional perspective approximately over 90 percent of employees execute processes as they are part of operations working for different B2B clients in different industries. Out of the remaining 10 percent approximately 1.000 employees are focused on solution design which includes the definition and optimisation of processes as well as IT, as all processes are at least supported by technology, if not fully digital.

\subsection{Relevance of process analytics and challenges}
As outlined the company is providing processes in a BPO model. Robust processes and their management are vital for successful business. Besides traditional forms of process management, process analytics provides multiple additional opportunities. This is especially true in case of outcome based pricing where the service provider is payed for a delivered outcome. While the price is stable service providers can reduce cost which immediately translates into financial gain. But also in case of input based pricing process analytics can help. While optimisations translate into lower chargeable hours client satisfaction can be increased and contract extension becomes more likely. 
The company leverages different channels ranging from voice (telephony) to non-voice (e.g. chat or social media). While the majority of interactions is processed by human agents emergent technologies such as chat bots augment human experts \citep{kraume2019customer}.
However, automation of processes can be expensive. Process analytics can help identify the best processes and thereby reduce cost of automation. A further advantage in a people intensive business is the utilisation of process mining techniques to find out about maverick processes and educate employees how to increase the quality of their work in the respective channel and especially when processes require an agent to work in multiple channels. 

Thus, many potential benefits exists for the company in applying process analytics. Yet, at the same time it became clear that successful application of process analytics goes beyond choosing the right technology and vendor. Various challenges arose which match with the research challenges outlined in chapter \ref{sec:pa}:

\begin{itemize}
    \item Process analytics and the subsequent automation of business processes allows to compete in a growing market. On an individual level, it enables employees to move on to more rewarding tasks fostering personal development given a growing market. However, individuals fear job loss which hinders process analytics efforts (see RQ-P \ref{rqp3}).
    \item Due to the corporate past and the decentralised organisation the company's entities had sometimes to compete and consequently differentiate against each other. Incorporating process analytics among this heterogeneous landscape raises many challenges, in particular towards achieving acceptance for new solutions and technologies (see RQ-G \ref{rqp2}).
    \item Many technological elements need to be considered when implementing data-driven approaches for BPM such as PBPM or process mining (see RQ-T \ref{rqt3}). To evaluate, purchase, combine, implement and manage the right technologies consistently many people with different backgrounds and degrees of data- and technology- and process-literacy need to collaborate. This does not work without joint understanding (see RQ-C \ref{rqc2}).
    
\end{itemize}

\subsection{Overcoming challenges via reference model}
\begin{figure}
    \centering
    \includegraphics[width=\textwidth]{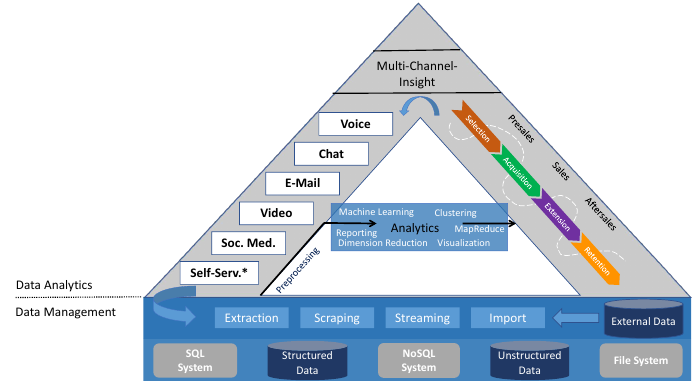}
    \caption{Reference Model: \emph{Analytics Framework} \citep{trautmann2017challenges}}
    \label{fig:arvato_ref_model}
\end{figure}

To address the presented challenges, representatives from different parts of the organisation invested time in an iterative process to align regarding the most important dimensions - in this case data, technology and process knowledge that all-together make up the \emph{Analytics Framework} \citep{trautmann2017challenges}. The process was time consuming primarily because it was difficult to anticipate all challenges that arise during evaluation, implementation and usage of process analytics to finally achieve business value. Many stakeholders were focused on technical aspects of processing large amounts of process data or automating business processes. Only after many iterations, the awareness for the importance of human and organisational concerns was raised. Finally, a reference model giving orientation and reducing complexity for all involved parties was achieved (see Figure \ref{fig:arvato_ref_model}).

While technologies such as ML are reflected in the middle the three main dimensions of the frame relate to the business context and allow different stakeholders to scope business challenges internally, but also externally with B2B clients when it comes to channel landscape (left), process landscape and customer journey (right) as well as data (bottom). Herewith all parties involved can handle complexities in a multi-channel environment more easily. While further dimensions such as industry specific use cases or language play a role these are not explicitly reflected. The value results from the fact that the reflected dimensions are agnostic of country, client and industry. Therefore they can be used widely across the 40.000 people organisation when designing or improving solutions. They also enable experts to leverage lessons learnt from other cases and tackle complex challenges and questions on time (see RQ-G \ref{rqp1}), software (see RQ-G \ref{rqp2}), privacy (see RQ-G \ref{rqc3}) and team formation (see RQ-G \ref{rqc4}) to name a few.     

\subsection{Matching the two frameworks}
As outlined earlier focus is often put on technology and despite its importance questions regarding expected benefits for and required contributions by people are insufficiently considered. Fig. \ref{fig:pa_concepts} changes this. While it is the starting point for further discussion it already now creates awareness regarding human and organisational concerns and the socio-technical perspective in general. 

The company initially struggled to implement process analytics given a very heterogeneous organisation spread over the globe and especially recent reorganisation. To select and implement the right technologies, a common understanding was important. Human and organisational concerns needed to be considered in. 
Fig. \ref{fig:arvato_ref_model} has been elaborated given the challenges of the company where business value is created across different channels and different phases of the customer life-cycle. Therefore the later two dimensions are made explicit besides data as a holistic perspective is crucial to foster business value. 
Reference models that take such a holistic perspective can support practitioners in making process analytics a success.

\section{Discussion}
\label{sec:discussion}

This paper proposes a definition and a conceptualisation of the term process analytics. The research objective is to raise awareness that techniques --- such as process mining --- are just one dimension of analytics \citep{holsapple2014unified}. While technologies based on process data are essential for deriving knowledge from data, using such technologies, e.g. for decision making in an organisation, poses various challenges \citep{Abbasi20161}. This paper's main contribution is to raise awareness for this aspect in the context of process analytics. It will hopefully stimulate a discussion on future research needs and the development of research streams within \ac{BPM}. 

A major implication for research concerns the evaluation of artefacts. The prevailing research paradigm in the field is \ac{DSR} which is used to design, instantiate and evaluate techniques and methods based on process data. The main objective of \ac{DSR} is to create useful artefacts \citep{Hevner2004Design}. 
The majority of publications in the field of process analytics evaluates the performance of their techniques and methods by benchmarking them against existing approaches with criteria such as accuracy (e.g. for predictions) or fitness (e.g. for process discovery). 
Such an evaluation is able to prove the \emph{goal attainment} of the artefact (in particular, its efficacy and validity) which is one crucial dimension of \ac{DSR} evaluation \citep{prat2015taxonomy}. However, in regards to evaluating whether an artefact reaches its goal further aspects to be considered are utility, feasibility and eventually also generality. Moreover, besides the goal another important dimension of artefact evaluation is the \emph{environment} which \citet{prat2015taxonomy} split into people, organisation and technology.

\citet{SACHAN2020113100explainableLoanUnderwriting} and \citet{stierle2020grm}, for instance, present case studies applying their techniques in a real-life setting at a company. This is a step in the right direction by providing a more substantial evaluation of the effectiveness of an artefact in a real situation. However, the artefacts are not actually embedded in the organisation to be able to evaluate environmental aspects. In particular, future work should employ \textquote{practice-based evaluation}\citep{prat2015taxonomy} and eventually user studies to evaluate aspects such as usefulness or ease of use which are crucial factors for technology acceptance \citep{davis1989perceived}. 
 
Last but not least, this work also contributes to practice. Practitioners can benefit from the proposed conceptualisation by gaining an improved understanding that successful process analytics goes beyond the right tool and technology. Further aspects need to be considered to gain value from process data in an organisation.


Besides these contributions, this work has several limitations. By building on existing work in the field of analytics and both case studies as well as interviews, the paper aims for an objective approach. Nevertheless, the choice of relevant dimensions is to some extent subjective. Furthermore, the publications listed in Table \ref{tab:concepts_analytics} are a selection from top IS journals and relevant outlets deemed most fitting within this paper's scope, but further references could be relevant. Another limitation is that all of the reviewed cases in Section \ref{sec:interviews} report from process mining usage, yet many other technologies for process analytics exist.

This work suggested various dimensions relevant to process analytics. Future work should discuss and evaluate the usefulness of the proposed concept of process analytics. In particular, the validity of this conceptualisation should be empirically evaluated  and each dimension's influence onto the value creation through process analytics should be assessed. Moreover, research should investigate the process of establishing process analytics in organisations, and, thus, evaluate the need for considering shifting importance of the dimensions in different stages \citep{vandewetering2019big}. Also, a structured literature review could be conducted that matches existing works to the various dimensions, and, in particular, to the various types of process analytics technologies as presented in Table \ref{tab:keyquestions_pa}. The results could be useful both for evaluating the proposed conceptualisation and for structuring the knowledge base on process analytics.
Last but not least, insights are worthless without action. \citet{trantopoulos2017external} find that \textquote{for enduring effects [of data analytics] on organisational performance external (and internal) data need to be molded into actionable insights that can be communicated throughout the firm}. Future work on process analytics should investigate the path from process insights and knowledge to action and ultimately to business value \citep{park2020general}.

\section{Conclusion}
\label{sec:conclusion}

The aim of this paper was to provide a better understanding of the term process analytics. For this purpose, ideas and concepts of the business analytics field were presented and transferred to the \ac{BPM} domain. Moreover, insights gained from interviews and publicly available case studies were used to develop and evaluate a conceptualisation of process analytics. 

With more and more organisations engaging in process analytics --- particularly through process mining --- many questions arise that research has to answer. Indeed, much of the existing work in the analytics field can serve as inspiration, but the specific nature of business processes --- and hence, process analytics --- call for a new research stream. Various scholars have already acknowledged this \citep{eggers2020turning,grisold2020adoption,mci/Brocke2021}, and hopefully, many more will follow.
 \section*{Declarations}
 The authors report there are no competing interests to declare.

\subsection*{Ethical Approval}
Not applicable
\subsection*{Funding}
No funding was received to assist with the preparation of this manuscript.
\subsection*{Availability of data and materials}
Not applicable
%
%
%

 \bibliography{references}
\appendix
\section{Interview Guide}
\label{appendix:sec:interview-guide}

\subsection{Process Mining in Enterprises}
{\it Respondents: Top management}
\begin{enumerate}
		\item What is Process Mining and why should managers care about it?
		\item Why did your organization decide to invest in Process Mining?
		\item How did your company approach the implementation of PM? Tell us about your organization's Process Mining journey.\\ 
		(e.g., duration, project management, trainings etc.)
	\end{enumerate}

{\it Respondents: Topic owners, IT}
\begin{enumerate}
	\item Could you describe what Process Mining is in 3-5 sentences to somebody who has never heard of it? 
	\item Why did your organization decide to invest in Process Mining? Did you have a particular goal? Did you substitute any existing system with PM?
	\item How did your company approach the implementation of PM? Tell us about your organization's Process Mining journey. 
\\	(e.g., duration, project management, training etc.) 
		
	\item Can you describe the different phases of the implementation?
			\begin{itemize}
				\item Who was involved in the process?
				\item How important was IT in the implementation process?
				\item What were the barriers? Challenges? Success factors?
			\end{itemize}
		
		\item What could have been done better during the implementation? What would you do differently next time?
\end{enumerate}

{\it Responders: End Users}
\begin{enumerate}
	\item Can you describe what Process Mining is in 3-5 sentences to somebody who has never heard of it? 
	\item Were you involved in the decision process of whether to implement Process Mining in your organization/department?
	\item How did your company approach the implementation of PM? 
\\
	\item How did your company approach the implementation of PM? Tell us about your organization's Process Mining journey. 
\\	(e.g., duration, project management, training etc.) 
		
	\item Can you describe the different phases of the implementation?
			\begin{itemize}
				\item Who was involved in the process?
				\item How important was IT in the implementation process?
				\item What were the barriers? Challenges? Success factors?
			\end{itemize}
		
\item What could have been done better during the implementation? What would you do differently next time?
	
\end{enumerate}

\subsection{Management and the Governance Structures}
{\it Respondents: Top management}
\begin{enumerate}
	\item Could you please describe how the organization covers costs for process mining implementation and use?	(i.e. the Business Operating Model)
	\item How did you communicate within the organization about the objectives, actions and results of Process Mining?
\end{enumerate}

{\it Respondents: Topic owners, IT}
\begin{enumerate}
	\item Could you please describe your organizational structure and how Process Mining is embedded in your organization? \\ (e.g., center of excellence, IT department, used and administrated by a business unit: centralized or decentralized)
	\item Could you please describe how the organization covers costs for process mining implementation and use? 	(i.e. the Business Operating Model)
	\item How do you support users of Process Mining in your organization?
	\item How did you communicate within the organization about the objectives, actions and results of Process Mining?
\end{enumerate}

{\it Responders: End Users}
\begin{enumerate}
	\item How does the management support you in using PM? Which resources were you provided with? Which support structures exists?
 \\	(e.g., training, communities, blogs, ...)
	\item What support structures would you like to have to use Process Mining in your organization better?
	\item How did the management communicate the objectives, actions and results of PM withinthe entire organization?
\end{enumerate}

\subsection{Benefits and Learnings}
{\it Respondents: Top Management, Topic Owners, IT}
\begin{enumerate}
	\item How does PM help your company to achieve its goals? How do you measure PM success?
	\item What has changed in your organization with PM? What can PM do for the organization what was not possible before?
	\item Knowing what you know today: Would you recommend your company to invest in process mining again?
\end{enumerate}

{\it Responders: End Users}
\begin{enumerate}
	\item How does your organization benefit from using process mining?
	\item What has changed in your organization with PM? What can PM do for the organization what was not possible before? 
	\item Knowing what you know today: Would you recommend your company to invest in process mining again?
	
\end{enumerate}
\end{document}